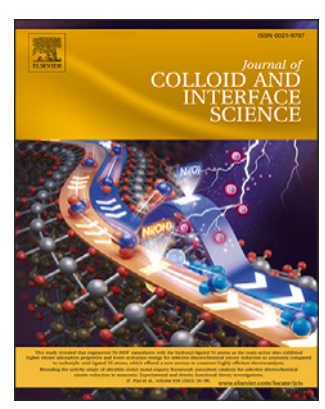

Regular Article

# Using wetting and ultrasonic waves to extract oil from oil/water mixtures

Yifan Li [a], Jesús. M. Marcos [b], Mark Fasano [c], Javier Diez [d], Linda J. Cummings [c], Lou Kondic [c], Ofer Manor [a,*]

[a] The Wolfson Faculty Department of Chemical Engineering, Technion — Israel Institute of Technology, Haifa 3200003, Israel
[b] Departamento de Física, Universidad de Extremadura, 06006 Badajoz, Spain
[c] Department of Mathematical Sciences, New Jersey Institute of Technology, Newark, NJ 07102, USA
[d] Instituto de Física Arroyo Seco, Universidad Nacional del Centro de la Provincia de Buenos Aires, and CIFICEN-CONICET-CICPBA, Pinto 399, 7000, Tandil, Argentina

## GRAPHICAL ABSTRACT

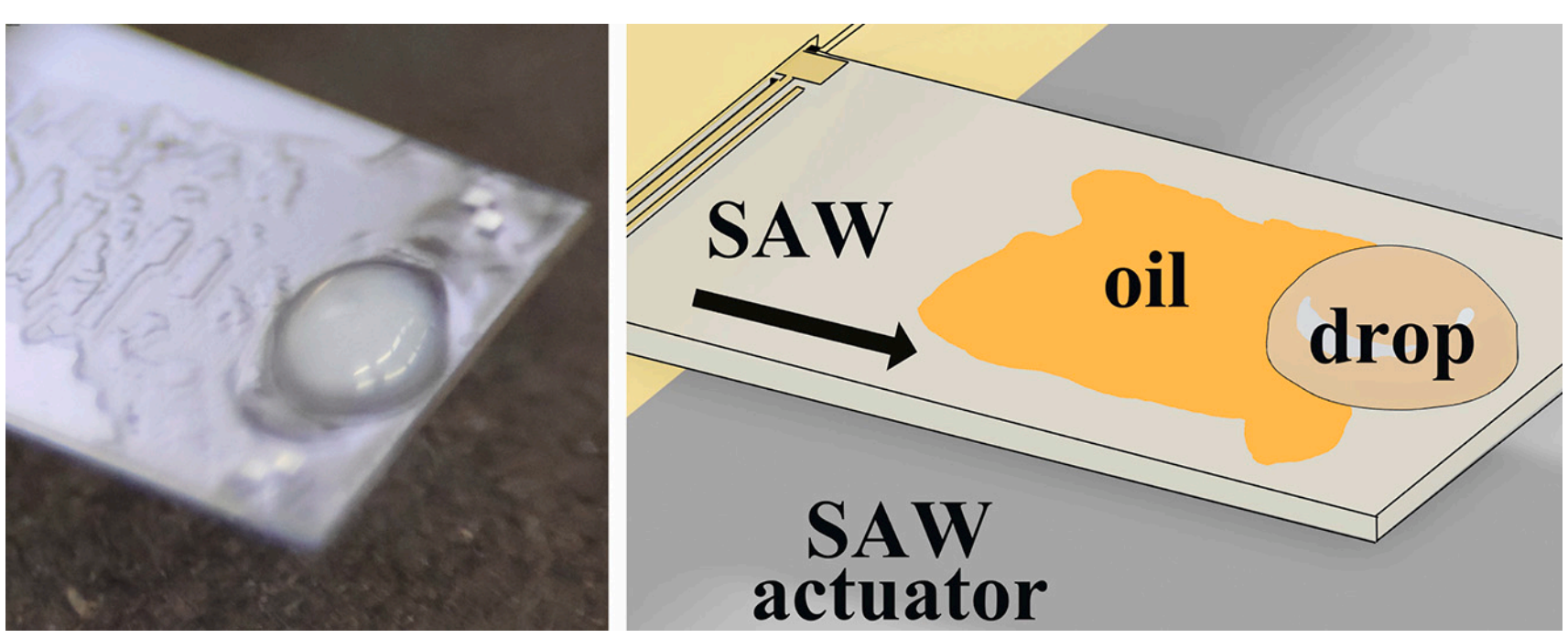

## ABSTRACT

Oil and water placed atop of a solid surface respond differently to a MHz-level surface acoustic wave (SAW) propagating in the solid due to their different surface wetting properties. We observe that, under SAW excitation, oil films, whether non-organic silicon oil or organic sunflower oil, are extracted continuously from sessile drops, comprising emulsions of the oil in question in a solution of water and surfactants. The mechanism responsible for the extraction of oil from the mixtures is the *acoustowetting* phenomenon: the low surface tension oil phase leaves the mixture in the form of 'fingers' that, away from the drop, spread opposite the path of the SAW. The high surface tension water phase remains at rest. Increasing either the SAW intensity or the oil content in the mixture enhances the rate at which oil leaves the emulsion. We further observe acoustic-capillary flow instabilities at the free surface of the oil film and the formation of spatial gradients in the emulsion oil-concentrations in the presence of SAW. Our study suggests the potential for using SAW for heterogeneous removal of oil from oil-in-water mixtures to complement current phase separation methods.

## 1. Introduction

One of the principal causes of shortages in drinking water is the use and pollution of water by industry, e.g., pharmaceutical and healthcare emulsion processing [1] and petroleum extraction, [2–4] oil extraction in the food industry [5,6], and domestic processes that produce gray water [7]. A key process when reclaiming water is the separation of oily liquids from the water body. The established techniques for separating oil from water require extensive energy investment and add-on chemicals. At the large scale, this means employing high-power distilla-tion [8] or chemicals to force coagulation/flocculation of oil droplets [9,10]. These methodologies have been used for approximately two hundred years and are compatible with large-scale water recovery facilities. Micro-scale setups for breaking oil-in-water emulsions down to their constituent phases are preferable to existing macro-scale emulsion-breaking methods for the local recovery of gray water in domestic housing and small-scale industrial applications. Here, we explore the interplay between capillary and acoustic stress in oil–in–water mixtures for heterogeneous phase separations.

* Corresponding author.
*E-mail address:* manoro@technion.ac.il (O. Manor).

It was recently discovered that MHz-frequency SAWs, traveling along the surface of a solid substrate, generate different responses in oil and water. Initial studies concentrated on silicone oil; in particular, Rezk et al. [11–13] excited a drop of silicone oil on the horizontal substrate of SAW actuators and observed that an oil film emanates from the drop and dynamically wets (spreads over) the solid in the direction opposing that of the traveling SAW, a phenomenon that has been called *acoustowetting*. The wetting rate of the oil films correlates with the oil viscosity; with the SAW normal particle-velocity amplitude at the solid surface (which is porportional to the square root of the SAW intensity), $\omega A$, where $A$ is the normal displacement amplitude induced by the SAW at the solid surface and $\omega$ is the angular frequency of the SAW; and with the film thickness, $H$, which was measured to be tens of micrometers in magnitude (the exact value is determined by a balance between acoustic and capillary stresses). Moreover, Collins et al. [14] and Manor et al. [13] suggest specific stable values for the film thickness $H$, that correlate with the wavelength of ultrasound leakage off the SAW. In particular, for a silicone oil of a 20 mN/m surface tension excited by a 20 MHz SAW, $H$ was measured to be $H \approx 20$ μm, regardless of the acoustic power [11,12].

Altshuler et al. [15,16] further considered the case of partially wetting (finite three-phase contact angle) solutions of water and surfactants. They show that the acoustowetting phenomenon is the product of a balance between capillary and acoustic stresses in the liquid film. Capillary stress, $\gamma/H$, resists the formation and spreading of the films on the solid, while acoustic stress, $\rho(\omega A)^2$, drives these effects, where $\gamma$ is the surface tension at the free film surface and $\rho$ is the liquid density. Altshuler et al. [15,16] suggest that the dynamic wetting (spreading) of liquid films under the action of SAW is governed by the parameter $\theta^3/\mathrm{We}$, where $\mathrm{We} \equiv \rho(\omega A)^2 H/\gamma$ is an 'acoustic' Weber number, which is scaled by the third power of the three phase contact angle between the liquid film, vapor, and solid substrate, $\theta$. This parameter measures the ratio between acoustic and capillary stresses in the liquid film. Where $\theta^3/\mathrm{We} > 1$, the film is governed by capillary stresses, resulting in a static drop. Where $\theta^3/\mathrm{We} < 1$, the film is governed by acoustic stresses, leading to dynamic wetting. The theory is compatible with previous silicone oil experiments, in which silicon oil supporting small contact angles at equilibrium ($\theta \approx 0$) was found to dynamically wet the solid in the range of SAW intensities studied. In the case of water films, Altshuler et al. [15,16] have confirmed experimentally the validity of the theoretical predictions, with a transition in the water film dynamics from standing still ($\theta^3/\mathrm{We} < 1$) to dynamic wetting ($\theta^3/\mathrm{We} > 1$).

Horesh et al. [17] further added gravity to disrupt the balance between acoustic and capillary stresses and emphasize the differences in oil and water/surfactant solution response to SAW. The analysis was found useful for measuring acoustic stresses in partially wetting liquids: the authors found that silicone oil films were able to leave a reservoir and continuously climb up the surface of a vertical SAW actuator against gravity. However, water/surfactant solutions under the same conditions left the reservoir and climbed to a finite height of just a few millimeters up the same vertical SAW actuator. The finite height to which water/surfactant solutions climb is determined by the competition between gravitational, capillary, and acoustic stresses in the climbing films. In a similar experimental setup, Satpathi et al. [18] investigated the shear-driven pinch-off of droplets from spatially periodic fluid protrusions in the film, which provides another example of interaction between acoustic and capillary stresses in liquid films.

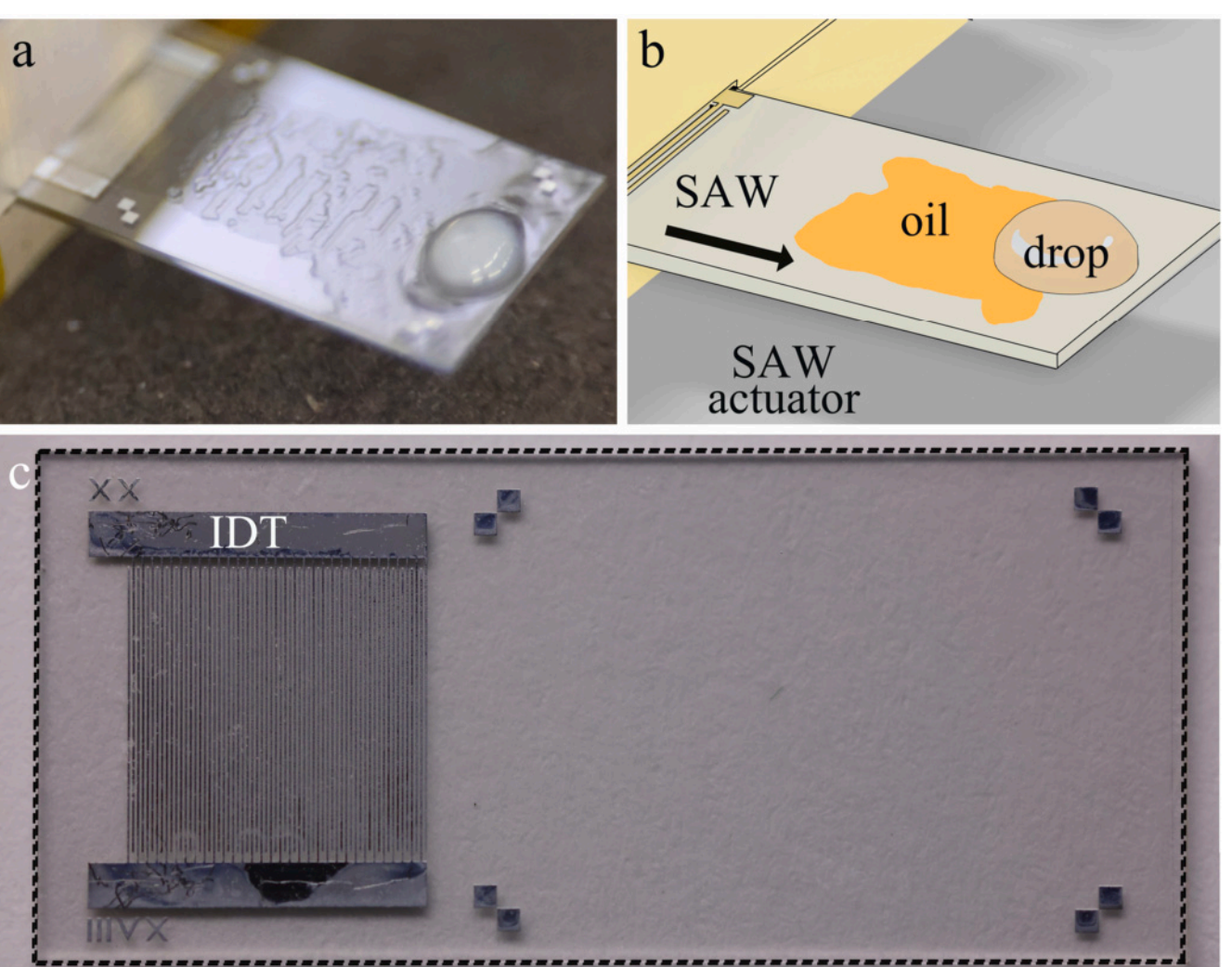

**Fig. 1.** (a) Our experimental setup (view from above) where we position the SAW actuator that supports an emulsion drop using a 3D printed plastic case that connects the actuator to power and (b) a schematic sketch (view from above) of the same system, further illustrating the oil film emerging from the emulsion sessile drop under SAW excitation. (c) The SAW actuator (dimensions ∼ 24.5 mm×10.8 mm×0.5 mm) is comprised of inter-digital metal electrodes fabricated atop a transparent piezoelectric lithium-niobate (LN) substrate (The local combination of electrodes and piezoelectric substrate are termed inter digitated transducer—IDT); the sides of the metal squares fabricated atop the LN substrate, away from the IDT, are 0.5 mm long.

In what follows, we use the differing responses of oil and water to SAW excitation to extract oil from a sessile drop of oil/water/surfactant mixtures. Silicone and Sunflower oils satisfy the inequality $\theta^3/\mathrm{We} \ll 1$ ($\theta \ll 1$ radians; $\gamma \sim 20$ mN/m and 30 mN/m for Silicon oil and for Sunflower oil, respectively; $H \sim 25$ μm; and $\rho \sim 1$ g cm$^{-3}$) in the presence of a SAW with intensity that corresponds to $\omega A \approx 1 - 1000$ mm/s, a range commonly used for SAW microfluidics; the SAW induced acoustic stress in the liquid governs oil film dynamics. However, the water/surfactant phase (with a much higher contact angle $\theta \approx \pi/6 - \pi/3$ radians ($30 - 60°$) and $\gamma = 40 - 70$ mN/m) requires significantly higher SAW intensity to render $\theta^3/\mathrm{We} < 1$ and thus produce spreading. We experimentally employ SAW intensities below such values, therefore we may expect a different response of oil compared to that of the water or water/surfactant mixture. Exploration of these different responses of the emulsion components is our main focus.

The rest of this paper is structured as follows. The section *Experiment* describes our experimental procedure. In the subsequent section, *Formation and dynamics of oil films*, we discuss the silicon oil film that leaves emulsion drops in our experiments. We further describe the temporal and spatial variations in the composition of the emulsion drops in *Variations in silicon oil content in the emulsion drops*, show the generality of the phenomenon by using edible sunflower oil instead of the silicon oil used in most of our experiments in *An emulsion of sunflower oil in water*, and give our further insights and perspective in *Conclusions*.

## 2. Experiment

We prepare surfactant—sodium dodecyl sulfate (SDS) and (separately) Tween 20—stabilized emulsions consisting of silicone oil droplets, approximately 230 nm in diameter (median drop size), in deionized water. Electrostatic and steric repulsion forces generated by the surfactants, adsorbed on the emulsion droplets, counteract attractive van der Waals forces and support the emulsion's kinetic stability. The drop size in the emulsion remains stable for 12 to 18 months when kept in a closed vessel on a lab shelf at approximately 20 °C. No coalescence or creaming appears during this time, and the size of the emulsion

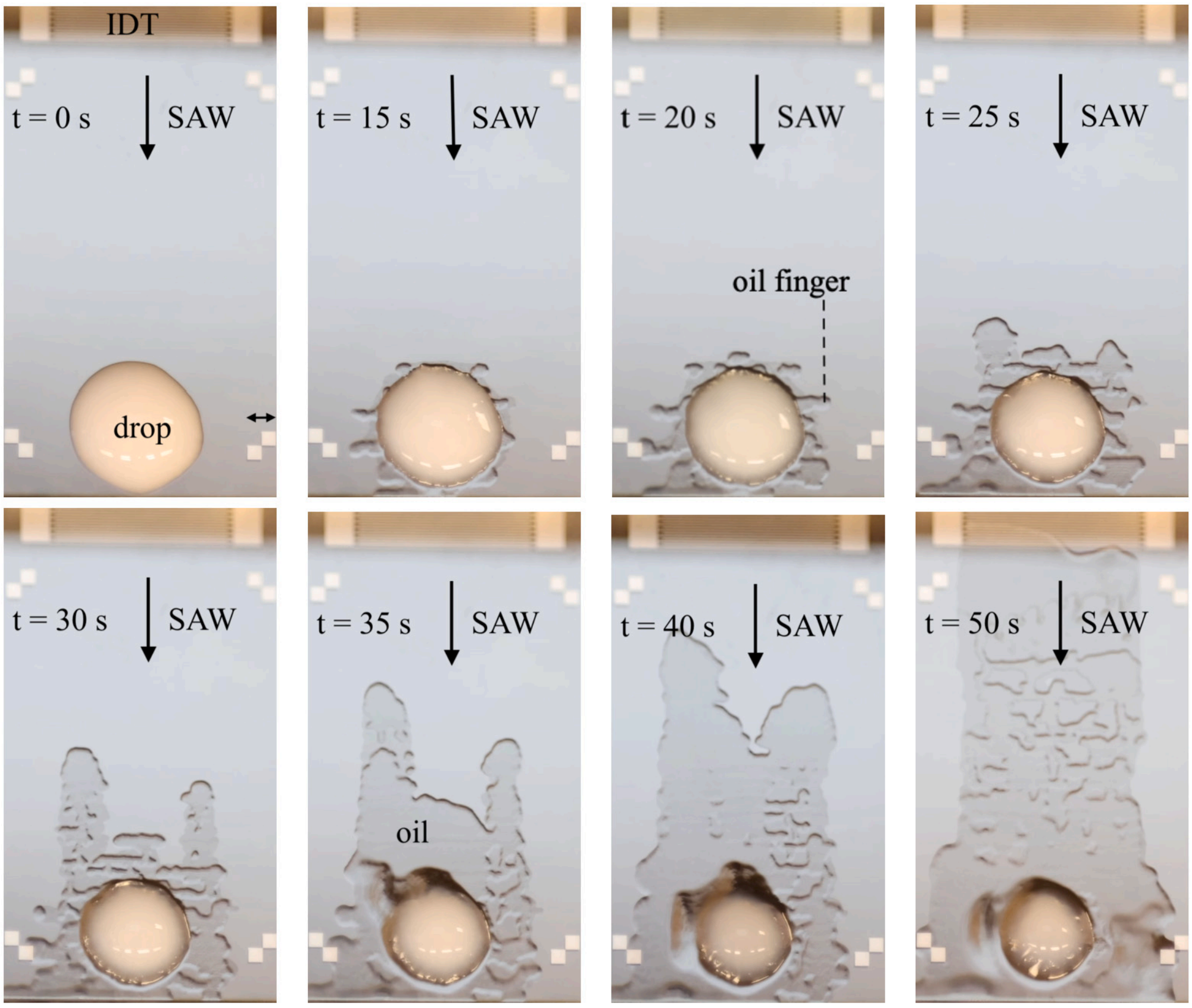

**Fig. 2.** Top view of a typical experiment; in this and the following figures we use a 10 µl emulsion (40% silicon oil in water emulsion, at 230 nm oil droplets diameter) drop at lab ambient conditions (50% humidity, 20 °C) if not specified differently. The SAW excitation amplitude is $A = 1.8$ nm. Time $t = 0$ corresponds to the moment we observe the appearance of oil at the drop circumference, which here occurs after a waiting time of $t_w = 190$ s after the commencement of SAW excitation. Initially ($t = 0 - 20$ s), we observe that (transparent) fingers of oil leave the drop transverse to the path of the SAW. After $t = 20$ s the oil fingers that have emerged from the drop change direction and spread in the direction opposite the SAW propagation. At yet later times, the film of oil spreading between the emulsion drop and the electrodes on top of the SAW actuator (IDT) develops surface patterns characterized by a length scale of $\approx 0.5$ mm. The double arrow in the $t = 0$ image is 1 mm long. Movies 1 and 2 in the Supplementary Information show both the top and side views of the experiment.

droplets remains approximately the same; see further details in Supporting Information [19].

We place a drop of oil-in-water emulsion atop the solid substrate of an acoustic actuator and introduce a propagating surface acoustic wave (SAW)—a Rayleigh wave of nanometer displacement amplitude—that travels in the solid substrate and comes in contact with the sessile drop. Fig. 1 shows our experiment. We use $\omega/2\pi = 20$ MHz-frequency actuators, which generate SAW of 200 µm wavelength in the substrate of the actuator, at power levels that generate normal displacement amplitude of $A = 0.5 - 2.5$ nm at the surface of the actuator. These values of $A$ correspond to a surface displacement velocity (particle velocity) of approximately $A\omega = 60 - 300$ mm/s in the direction normal to the solid surface. The SAW introduces excess acoustic stress in the neighboring fluid, with the consequence of extracting oil from the emulsion drop, as discussed next.

The SAW frequency is similar to that used in previous studies, where oil [11,12] and water [15,16] films were successfully actuated by SAW. Once an oil film has left the emulsion drop and dynamically wet (spread over) the solid surface of the actuator, further oil release becomes more difficult, because now, to reach the emulsion drop and force more oil mass release, the SAW must travel through the solid substrate of the actuator under an oil film that enhances the SAW attenuation. The SAW attenuates under the film due to the diffraction of ultrasonic waves into the overlying oil. In our experiments, we minimize this attenuation, which increases with increasing SAW frequency, by using 20 MHz SAW, which is the lowest SAW frequency that support a stable SAW (Rayleigh wave) when using the 0.5 mm thick commercial piezoelectric lithium niobate wafers that are designed to support SAWs.

During the experiment, it is possible to distinguish between the emulsion and oil phases by color and wetting properties: the opaque yellow/white color of the emulsion is a consequence of the diffraction of light by the emulsion nano-droplets. A transparent, clear, liquid suggests a pure or near-pure phase, either oil or water. At equilibrium, water and water/surfactant solutions on a lithium-niobate surface support a finite three-phase contact angle of [15,16] $30 - 60°$, while the low surface tension silicon oil supports at equilibrium a vanishing contact angle on the same substrate. Hence, a pure water/surfactant phase appears in the form of clear/milky drops, and a pure oil phase appears in the form of a clear liquid film, identifying the fingers that leave the drop as silicone oil.

Fig. 2 shows typical experimental results. The first three images in Fig. 2 at $t = 0 - 20$ s show the initial time interval during which oil leaves the emulsion drop under SAW excitation. We observe an accumulation of transparent oil along the drop's rim, followed by fingers of oil leaving the drop transverse to the path of the SAW (e.g., see the image at $t = 20$ s) and spreading over the solid substrate. Oil fingers further form toward the back side of the drop, far from the source of the SAW (IDT) as well. However, the oil film—free from the emulsion drop—is transported toward the IDT, opposing the path of the SAW. Therefore, these oil fingers are not as clearly visible as the ones emerging from the sides of the drop, transverse to the path of the SAW.

We expect that the oil finger formation (rather than the spreading of a continuous film) may be due to a combination of the following two ef-fects: (i) a limited availability of oil near the contact line of the drops, similar to the vegetation-patch pattern formation that results from limited water sources [20]; and (ii) capillary instability transverse to the motion of the film, which may break the continuous film to isolated fingers of oil [21].

The initial evolution of oil fingers emerging from the emulsion drop is followed at $t = 25 - 50$ s (the remaining images in Fig. 2) by a continuous oil film surrounding the drop and spreading over the solid substrate in the direction opposing the SAW. Moreover, for $t > 25$ s, the oil film away from the drop shows surface patterns associated with variations in the oil film thickness. Specifically, the film thickness undulates, forming hills and valleys, which are characterized by the in-plane length scale of approximately 0.5 mm; while we do not have at this point a good understanding of the source of such a length scale, we note that it is different from the 200 μm wavelength of the SAW.

## 3. Formation and dynamics of oil films

Fig. 3 shows a side view of the emulsion drop, highlighting the initial formation of an oil film on top of the emulsion drop, which appears to be the source of the oil leaving the drop. Fig. 4 compares an experiment exposed to the laboratory atmosphere of approximately 50% humidity and one in a humidity chamber that provides approximately 85% humidity; as we will see, quantifying the effect of humidity helps us understand the mechanism leading to the formation of the oil film. The figure shows that the time taken for oil to leave the drop depends on the ambient humidity level, with the oil film emerging faster from emulsion drops at lower humidity, suggesting that evaporation plays a role.

Fig. 5 further explores the effect of humidity as the intensity of the SAW is varied. We observe the general trend that the 'waiting time' $t_w$, (the time period between the start of SAW and the first appearance of the oil film) is shorter if humidity is lower. Consistent results were found when using different surfactants (as stated, two types were used: charged Sodium Dodecyl Sulfate (SDS) or non-charged (oily) polysorbate-type nonionic surfactant which includes 20 repeating units of polyethylene glycol (Tween-20)) to stabilize the emulsion. Similar formation of oil films during the evaporation of the water phase of oil/water mixtures is described elsewhere [22,23]. It appears that the evaporation of the volatile water phase in the presence of the non-volatile oil phase renders the latter evermore concentrated in time near the free surface of the emulsion until it forms a continuous film. The oil film at the surface of the emulsion drop leaves the drop under the action of the SAW. Fig. 5 is also consistent with the expectation that an increase in humidity leads to a reduction in the water evaporation rate, and consequently an increase in the time required to form an oil film that could leave the emulsion drop. Thus, both the rate of water evaporation and the level of SAW intensity influence the waiting time, $t_w$, at which oil leaves the emulsion drop.

Once oil films leave the drop, they spread on top of the solid substrate in the direction opposite that of the SAW. This motion suggests the *acoustowetting* effect, known to lead to such behavior in the case of oil films that originate from one-phase oil reservoirs, e.g., oil drops [11–13]. In such one-phase oil systems the SAW leads to the formation and spreading of oil films of a thickness smaller than half the wavelength of the leaked ultrasonic waves that diffract off the SAW in the underlying solid substrate and propagate in the overlying fluid.

We employ 20 MHz-frequency SAW in our experiment, which leaks the same frequency ultrasound waves of approximately 80 μm wavelength. Hence, to verify our assertion about the acoustowetting phenomenon in our experiments, we proceed by measuring and discussing the thickness of the extracted oil film and show that it is indeed of a thickness smaller than half the ultrasound wave leakage off the SAW. Fig. 6 shows a side view image of our laser interferometry experiment [24–26], comprising a red laser source, mounted on a protractor to identify the angle of the laser relative to the oil film, and a camera mounted at the opposite angle to the laser source. The interaction of the laser with a nearly flat and smooth oil film results in light fringes of equal chromatic order (FECO). The light fringes appear as dark and bright areas that map spatial variations in film thickness. We estimate the height difference associated with a path along the surface of the film by crossing from bright to dark and then back to bright fringes or vice versa, i.e., along one period of fringes, using the expression

$$\Delta H = \lambda_0/(2n_3\cos(\theta_3)),\ \text{where}\ \theta_3 = \sin^{-1}(\sin(\theta_1)/n_3), \qquad (1)$$

with $\lambda_0$ the wavelength of the incident light, $n_3$ the refractive index of the oil, and $\theta_1$ the angle between the laser beam and the normal to the free surface of the oil film.

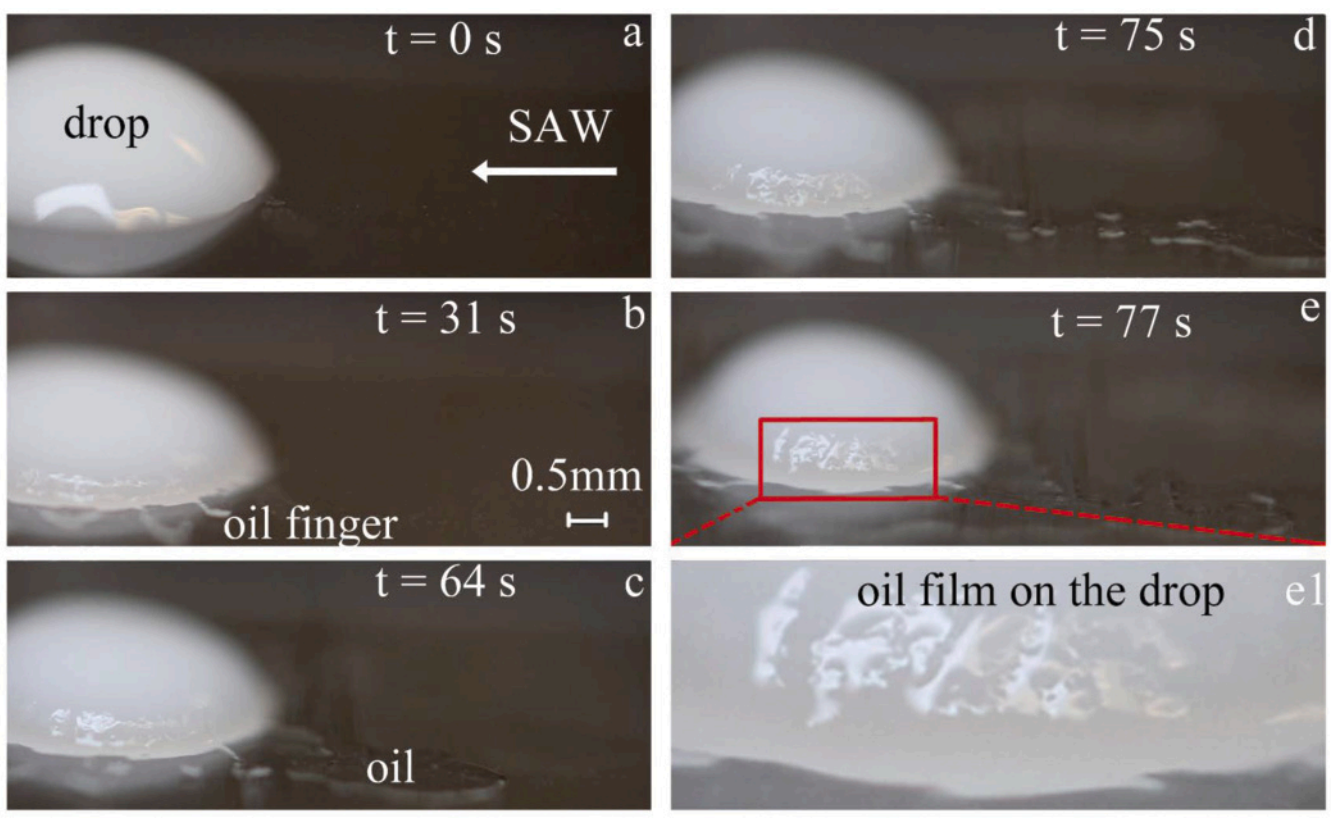

**Fig. 3.** Side view time-lapse of an emulsion drop under the influence of SAW, where $t = 0$ gives the time at which oil appears at the drop circumference; the SAW excitation amplitude is $A = 1.8$ nm. (a) Here, time $t = 0$ corresponds to $t_w = 150$ s from the commencement of SAW excitation. We observe the accumulation of an oil film atop the drop surface and along its rim. (b) shows the initial formation of oil fingers leaving the drop; these oil fingers become longer as time progresses (as shown in (c–e); the fingers merge at longer times to form a continuous oil film that spreads on the solid in the direction opposite to the SAW. See video #2 in Supplementary Information (SI).

Fig. 7 shows an example of an oil film extracted from a sessile emulsion drop (view from above). Here, the emulsion drop is located at the bottom left of the images in the top panel, outside the red spot of the laser, which is aimed at an oil 'finger' (see Fig. 2) emerging from the drop. The oil film that leaves the drop appears in the form of a bright-red and dark pattern. In the presence of SAW, we do not observe the usual regular light fringe (FECO) patterns that portray thickness variations along a film of smooth surface. Rather, the patterns, shown in Fig. 7(A), suggest that in the presence of SAW the oil films undergo abrupt spatial periodic changes in thickness, shown as cell-like spatial variations in bright and dark patterns (see A1, the zoom-in of Part A of Fig. 7). The appearance of these cell-like thickness variations in the presence of the laser beam suggests spatial variations in film thickness that correspond to 1/4 to 1/2 of the 635 nm laser beam wavelength, i.e., thickness variations of approximately 0.15 to 0.3 μm. These thickness variations are likely a consequence of a balance between the two most dominant mechanisms in this problem, capillary and acoustic stresses.

The cell-like patterns hinder the formation of ordered FECO patterns, and hence the measurement of the film thickness. To measure the film thickness, we turn the SAW off. Fig. 7(B), taken 1 second from the moment the SAW is turned off, shows that the film surface is sufficiently smooth at this time to support a well-ordered FECO pattern that allows for film thickness measurement. We find that the oil film is comprised of thick and thin steppe-like regions, illustrated schematically in Fig. 7(C). The thick parts of the film show the cell-like pattern in the presence of SAW. The thin parts do not support the cell-like pattern, showing well-ordered FECO patterns.

The thin parts of the oil film appear to be wetting oil layers of a typical thickness of 1-3 μm; see Methods for more details. The thicker part of the film is measured to be approximately 25 μm thick, between 1/4 to 1/2 of the ultrasound leakage wavelength of 75 μm (the ultrasound is of 20 MHz frequency and propagates at a phase velocity of approximately 1,500 m/s in oil). Hence, our observations are consistent with the film thickness anticipated in acoustowetting, suggesting that this is the mechanism that drags the thick oil films out of emulsion drops in the presence of the SAW.

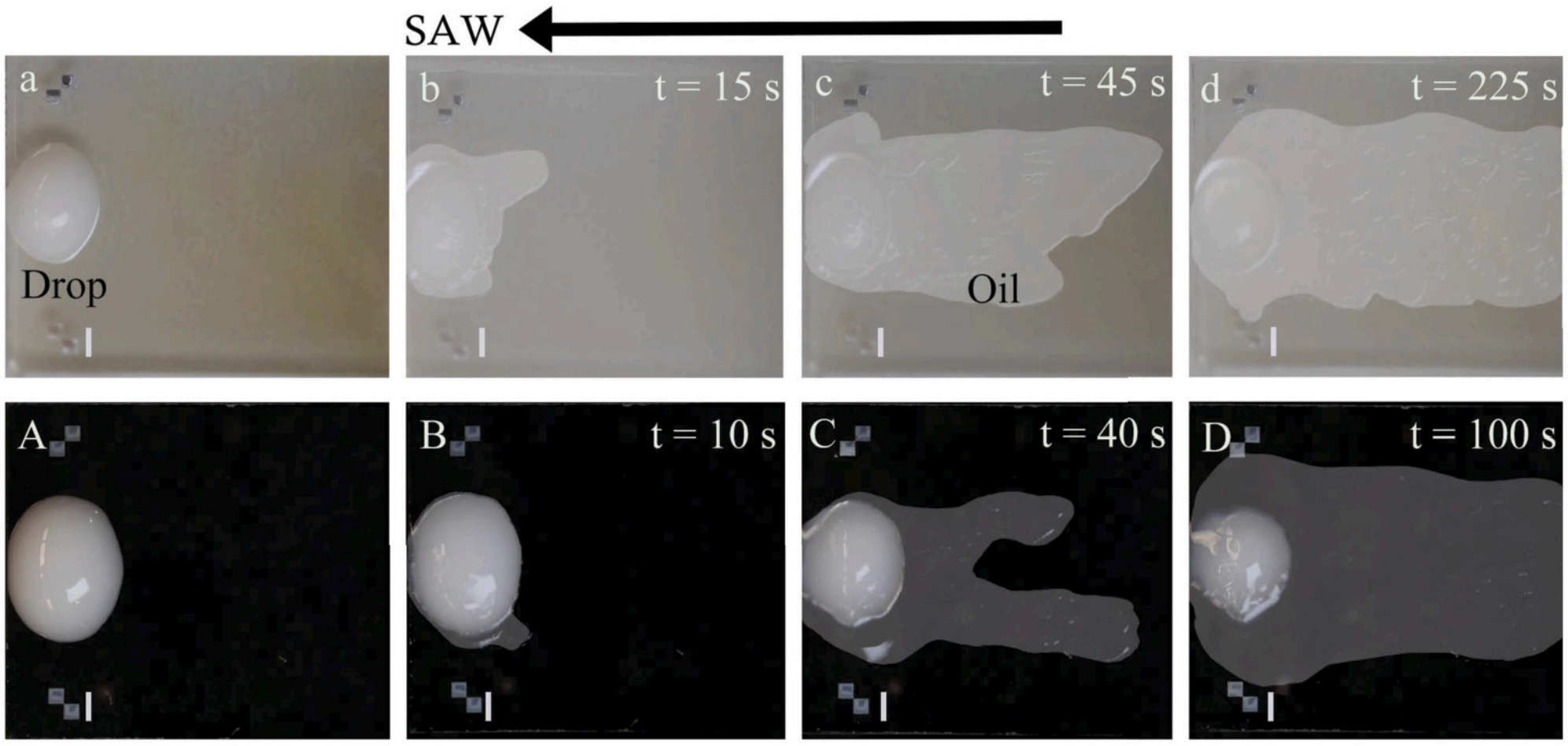

**Fig. 4.** Top view of a 10 μl emulsion drop under SAW excitation ($A = 1.5$ nm). (a-d) A drop is placed in a humidity cell (85% humidity) and (A-D) a drop is placed under lab ambient conditions (50% humidity). The time $t = 0$ is taken at the moment ($t_w$) at which we observe the presence of transparent oil at the circumference of the drop; in (a) $t_w = 465$ s and in (A) $t_w = 170$ s. We observe, both in the humidity cell and under lab conditions, that the oil films leave the drop and spread over the solid substrate in the direction opposite the path of the SAW (see the arrow that illustrates the direction of the SAW above the images), but at different times. The white scale line is 1 mm long. The difference in shade between the two experiments results from the presence of a transparent solid substrate between the drop in the humidity cell and the camera in (a-d), and the slightly non–circular perimeter of the drop takes place when the drop is in contact with the edge of the actuator on the left hand side of the images. See SI for additional figures and videos.

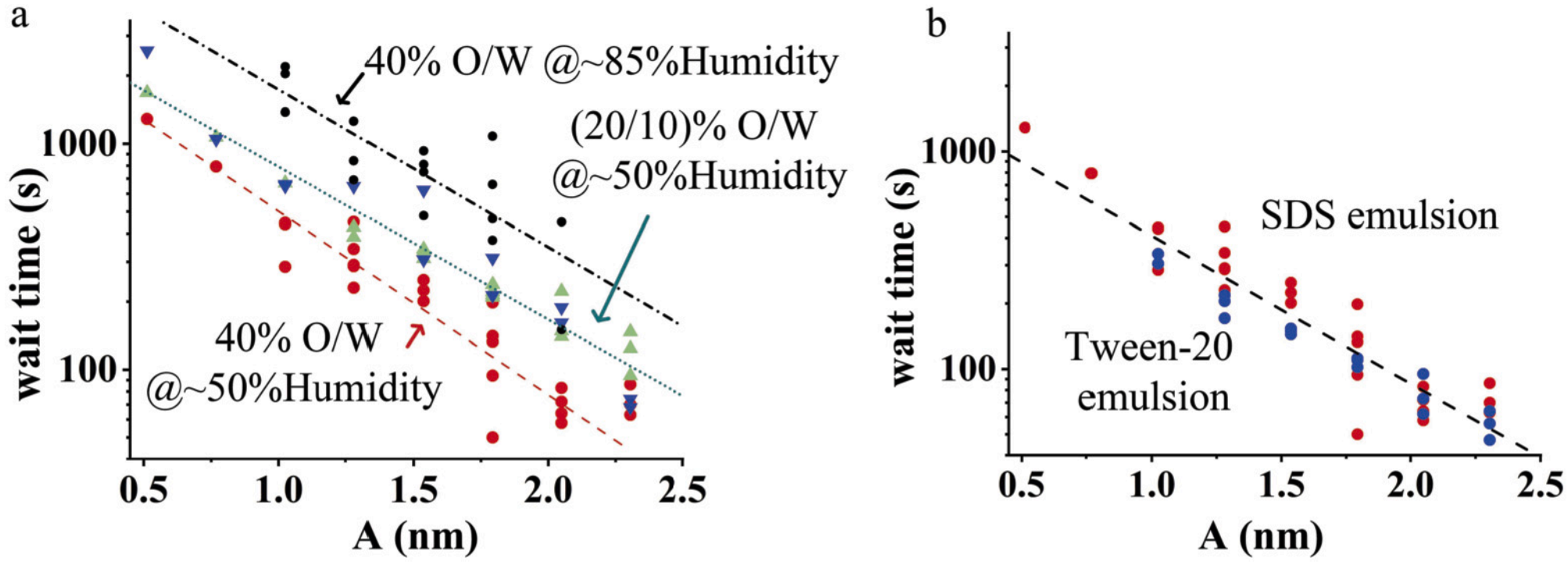

**Fig. 5.** Influence of SAW amplitude $A$ on the waiting time ($t_w$) for the appearance of oil films, for (a) different initial oil concentrations and ambient humidity levels, and (b) different surfactants. The symbols are measured points and the lines are a guide for the eye. The results for 10% and 20% oil are indistinguishable and as such we draw the same line for both, highlighted by (20/10)% O/W @ ∼ 50% Humidity. Every point stands for a single experiment; the waiting time may have of errors of ± 3 seconds by choosing the key-frames manually.

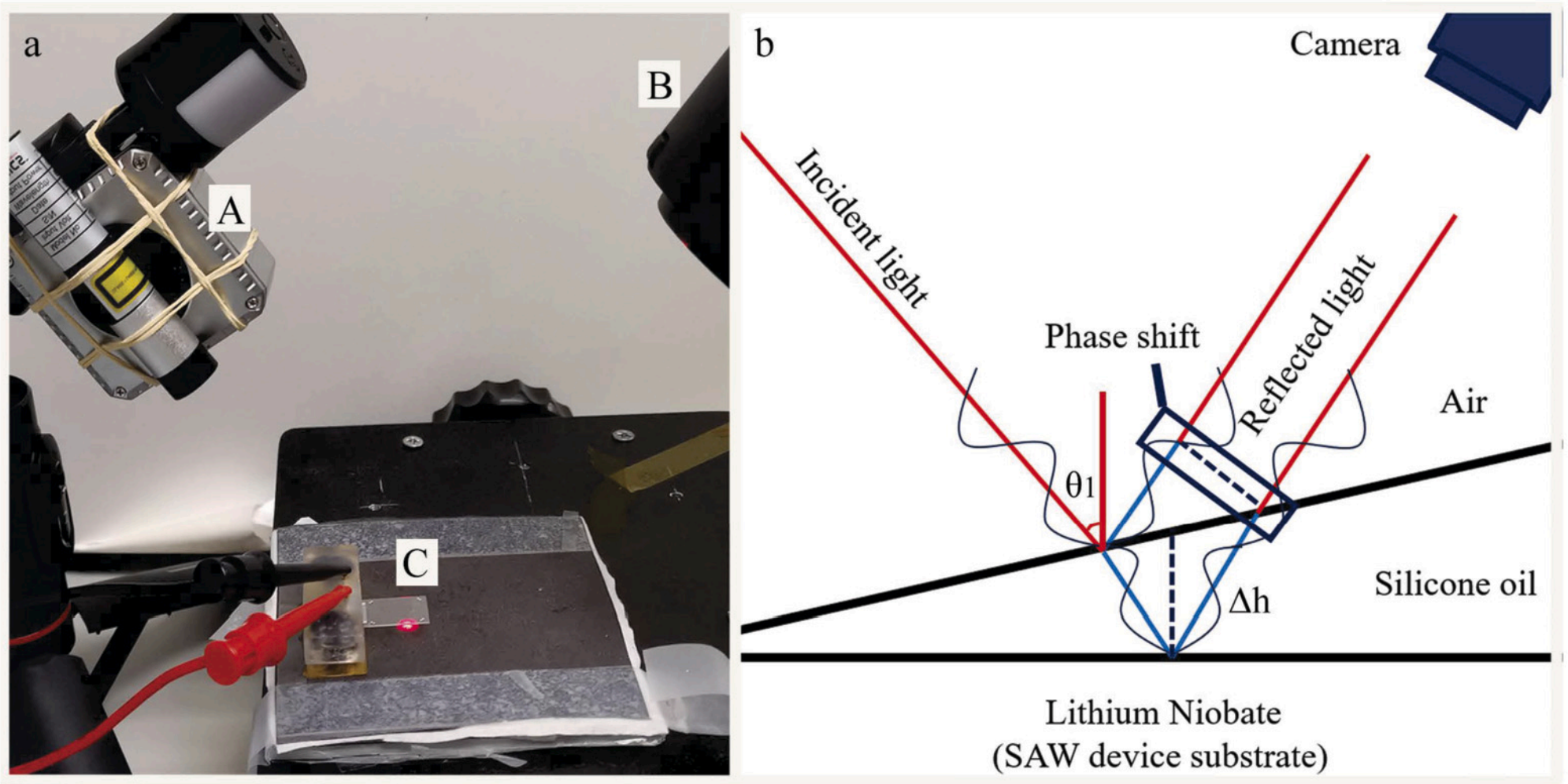

**Fig. 6.** (a) A side view image of our laser interferometry experiment, where the components marked by 'A', 'B', and 'C' are a digital protractor to monitor light angle emission with respect to the target surface, the position of the measured oil film, and a camera mounted at the opposite angle to the protractor, respectively. In (b), we illustrate the laser path in air and the oil film.

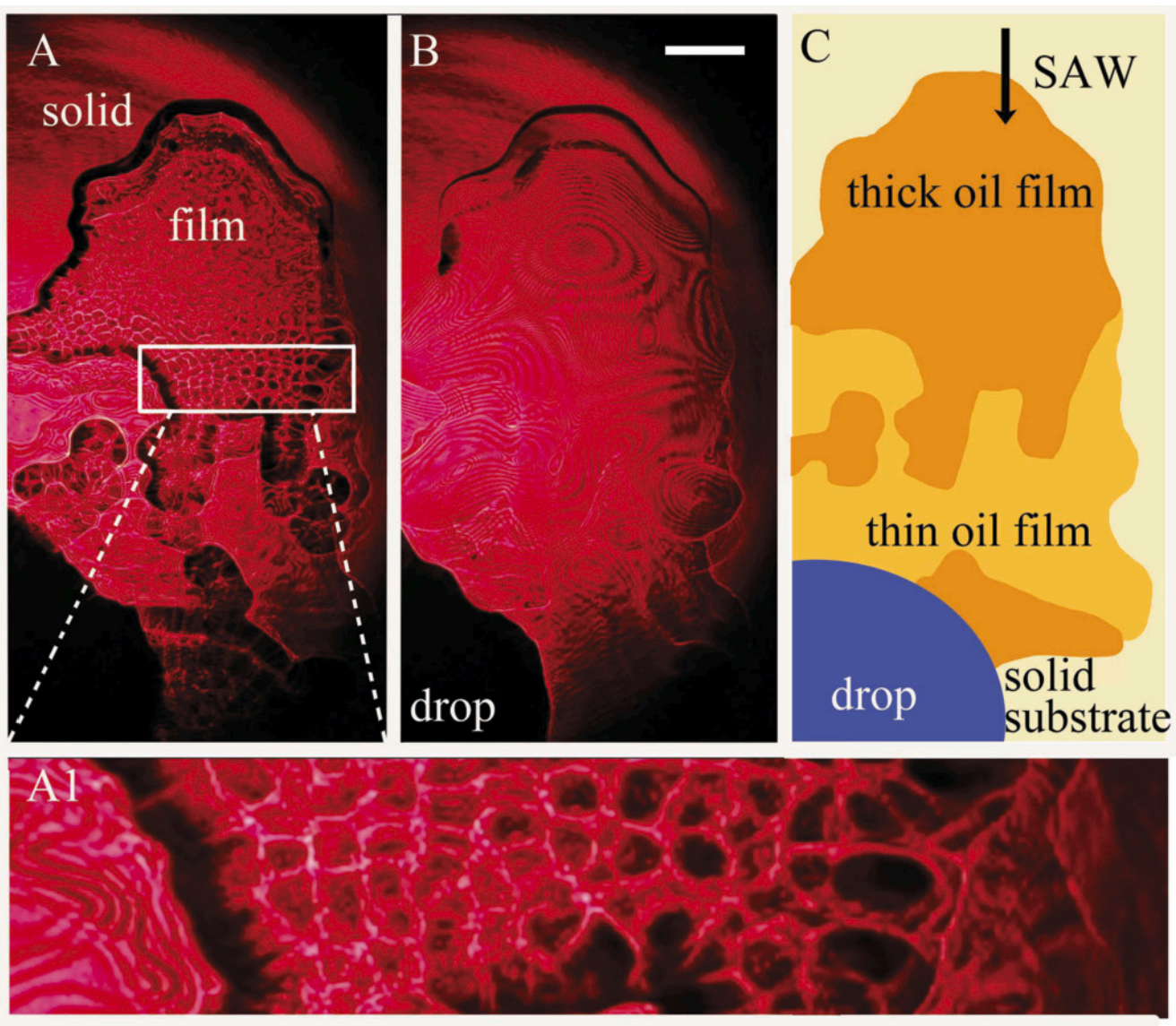

**Fig. 7.** Laser interferometry FECO (fringes of equal chromatic order) patterns on an oil film extracted from an emulsion drop (dark shadow at the left bottom corner) under SAW of amplitude $A = 1.5$ nm, where we show (A) the cell-like pattern while the SAW is active (image A1 shows zoom-in of the region bounded by the white rectangle in (A)), and (B) the orderly FECO patterns 1 s from the moment the SAW was turned off (the white scale bar is 0.5 mm long). Part (C) shows a sketch of (A) to help the interpretation of the results; the darker yellow color indicates a thicker oil film. (For interpretation of the colors in the figure(s), the reader is referred to the web version of this article.)

## 4. Variations in oil content in the emulsion drops

### 4.1. Spatial variations

One of the mechanisms that appears to contribute to the rate at which oil films leave the emulsion drop is the spatial distribution of oil emulsion droplets in the sessile drop under SAW excitation. Studies of the acoustowetting phenomenon [11,12,15,16] show that in single-phase systems, i.e., oil or water/surfactant solutions, films of liquid leave reservoirs of the pure liquids en-masse, in the direction opposite to the SAW propagation. However, in Fig. 2, we see that when the reservoir is an emulsion rather than a pure phase, the oil phase leaves the emulsion drop in the direction transverse to the path of the SAW. Once the oil film leaves the drop, it changes spreading direction and propagates in a direction opposite the SAW path. The initial motion of the oil film transverse to the path of the SAW suggests there is no oil in the front part of the drop. That is, there is a depletion of oil content in the front of the drop closest to the IDT (the origin of the SAW) and, thus, an inhomogeneous distribution of oil emulsion droplets in the sessile drop.

To measure the spatial distribution of oil emulsion droplets, we use emulsions characterized by different oil content and vary the intensity level of SAW excitation. We identify spatial variations in the concentration of the oil phase in the emulsion drop by monitoring the diffraction of white light therein. Using the principles of the Beer-Lambert law [27], we estimate the emulsion content during our oil extraction experiment using variations in the brightness level of the measured drop: as light passes through the emulsion, it is scattered by the oil droplets. Light scatter increases with the concentration of oil emulsion droplets and with the light path length as it travels through the drop.

Fig. 8 shows that the SAW renders the front of the drop (toward the source of the SAW) transparent (appearing black due to the color of the substrate below the emulsion drop), and the back of the drop opaque white. Contributions of the oil film (that forms on the drop) to light diffraction are assumed to be small. Thus, Fig. 8 suggests that the SAW 'pushes' the oil droplets to the back of the sessile emulsion drop. The transparent (dark) areas of the sessile drops indicate local areas of low oil concentration, where light undergoes weak diffraction. The non-transparent (bright) areas of the drops contain larger concentrations of oil droplets, where light undergoes appreciable diffraction. We note that this is only a rough map of oil concentration since both the curvature of the free surface, and the drop thickness, which also contribute to the extent of light diffraction, vary from location to location. Still, while our analysis gives only qualitative insight into the spatial distribution of the oil emulsion droplets in our experiments in real-time, it is sufficient to demonstrate the spatial non-uniformity of the oil droplet content.

In an effort to make this spatial nonuniformity of the oil droplet concentration in the emulsion drops easier to grasp, we coarse-grain the emulsion drop images to binary (white and black) color maps to represent areas of excess and depleted oil droplets therein. We commence this analysis by converting each drop image to a gray-scale scheme, where each pixel is represented by a range of numerical values between 0 and 1. Bright parts of the emulsion drops, represented by gray-scale numerical pixel values of 0.5 or larger indicate excess oil droplet concentration in the emulsion drop (relative to its initial oil droplet content) and are assigned the binary numerical value 1. Dark parts of the emulsion drops, represented by gray-scale numerical pixel values below 0.5, indicate areas of depleted oil droplet concentration in the emulsion drop and are assigned the binary value 0. We plot the ratio, Θ, of the emulsion drop bright area to the total area of the emulsion drop in Fig. 9.

An emulsion drop containing a uniform oil droplet concentration throughout its volume gives $\Theta = 1$, while a smaller Θ-value reflects that a portion of the drop volume is oil-depleted. We find that (1) the Θ-value is smaller under SAW excitation when compared to the uniform distribution ($\Theta = 1$), and (2) within the measurement uncertainty (deviation between 2 to 3 repetitions of the measurements), Θ is largely independent of the specific SAW intensity. Moreover, the change in Θ in the presence of SAW is sensitive to the overall concentration of oil in the drop. The change in Θ decreases when the initial oil content is increased, as may be seen by comparing Figs. 9(a) (lower oil concentrations) and (b) (higher oil concentrations); note the different vertical scales on the two plots. The fact that the SAW pushes the emulsion oil droplets along its path is a well-known phenomenon that originates from the radiation pressure the SAW inflicts on the emulsion droplets, see, e.g., studies by King [28], Shilton et al. [29], Li et al. [30], Rogers et al. [31].

To conclude, as a consequence of SAW pushing the oil droplets back within the emulsion drop, the oil intially leaves the drop from the sides and back, where the presence of increased oil content serves as a reservoir.

### 4.2. Temporal variations

To obtain a quantitative measurement of the oil content in the emulsion drops under SAW excitation, we monitor time variations in the diffraction of white light from above over a small area of $0.5 \times 0.5$ mm$^2$ about the drop apex to minimize light diffraction at the free surface of the drop. In this surface region, the drop surface is nearly flat, and is perpendicular to the path of the light beam. It is easy to identify variations in the local drop thickness (distance between the solid substrate and the drop apex), which is the length of the light path through the drop in our experiment. The measurement is conducted by monitoring the level of light diffraction between the center (apex) of the drops and the solid substrate below, along a light path of length $d(t)$ (twice the height of the drop's apex), serving as a proxy for the local concentration of oil droplets. We further measure the time variations of the drop height to identify the small variations in the magnitude of $d(t)$ during the measurement.

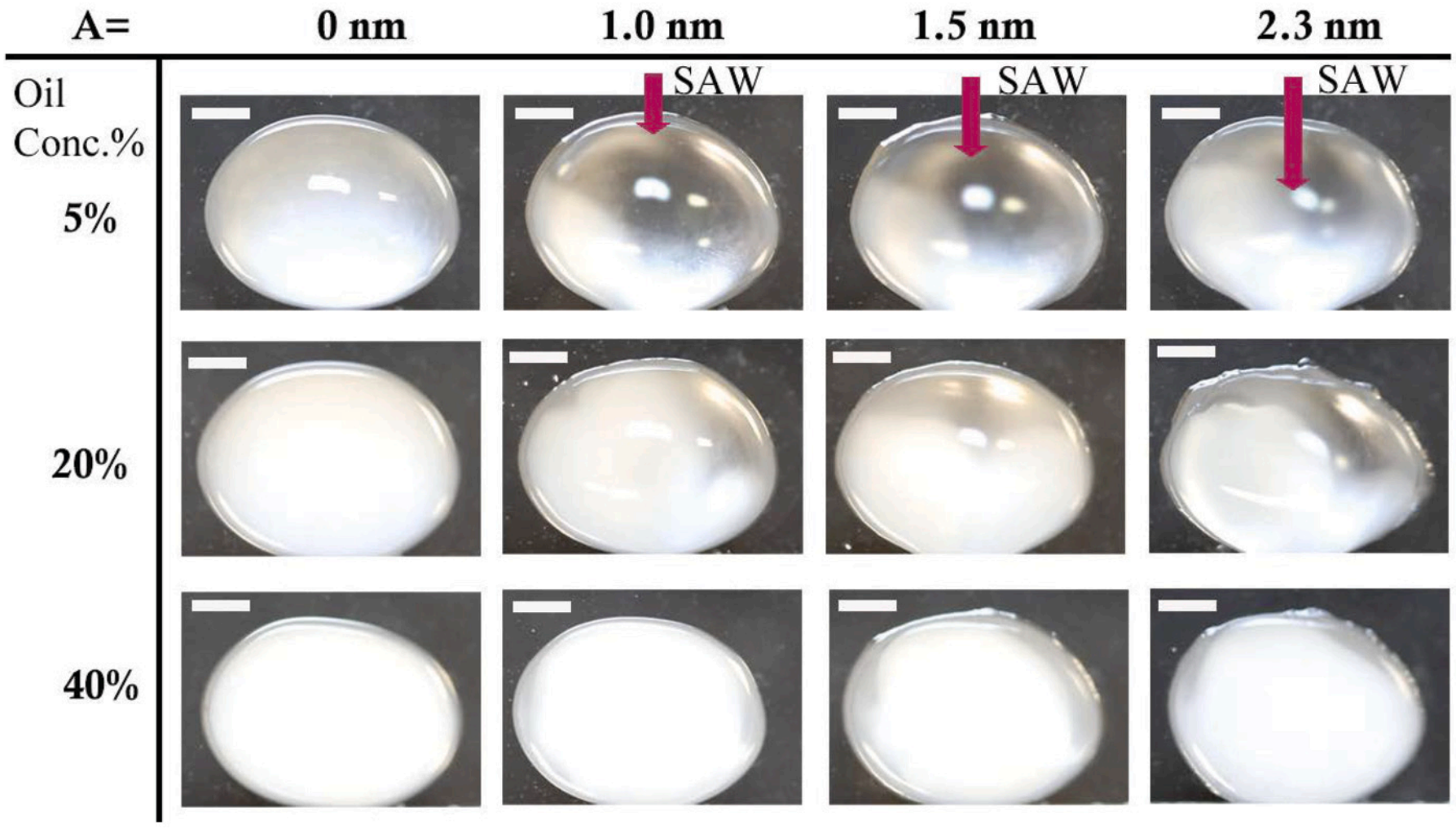

**Fig. 8.** Variations in SAW amplitude ($A$) and initial oil content in our sessile emulsion drops result in different spatial distributions of oil content under SAW excitation, indicated by transparent (dark) and non-transparent (bright) portions of the drop, which contain small and large concentrations of oil droplets, respectively. In this particular experiment, the size of the oil droplets in the emulsion was measured to be $4.7 \pm 0.5$ μm. The length of the scalebar in the images is 1 mm. See movies 4 and 5 in the SI as examples of the experiments.

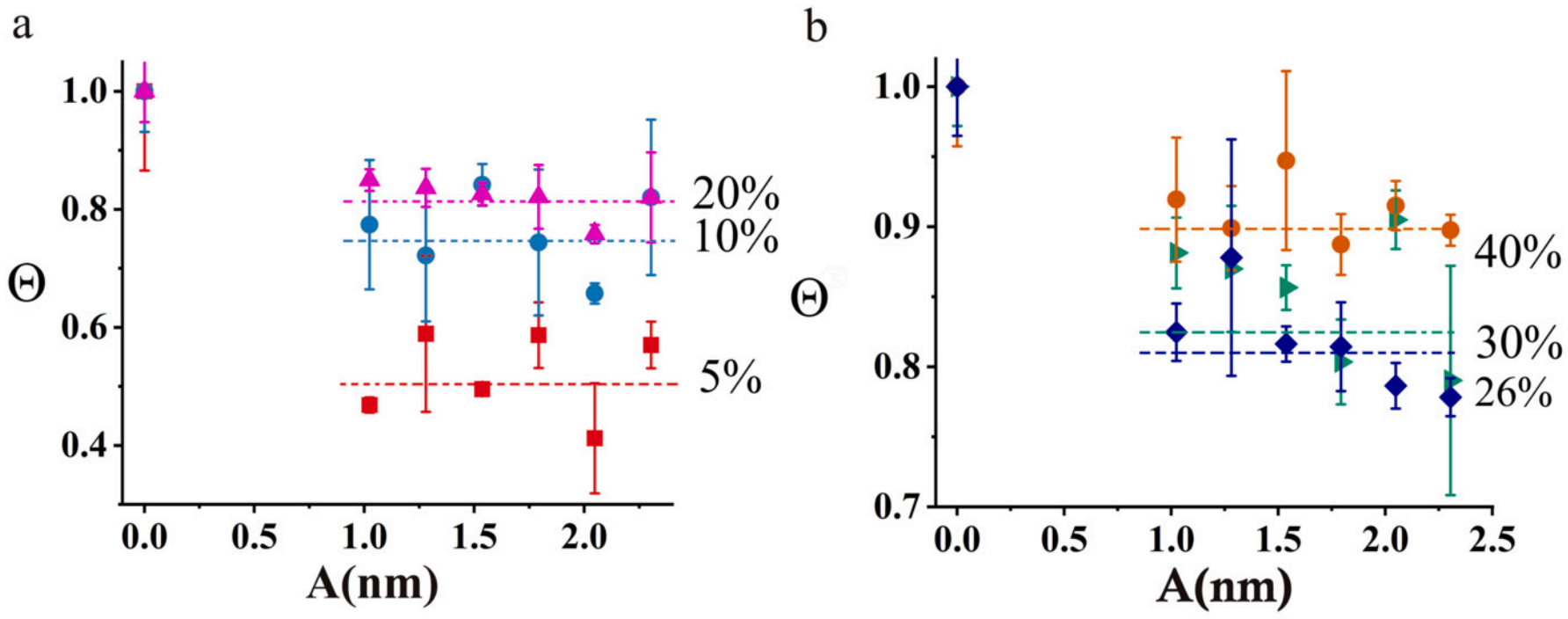

**Fig. 9.** Variations of Θ (the ratio between the area of the drop where oil concentration is in excess, to the full area of the drop, when viewed from above) with SAW amplitude $A$. The dashed lines represent the average value of Θ for three different measured values (with minima/maxima shown as errors) of the oil concentration (the corresponding experimental data points are matched by color and symbol). Part (a) shows results for emulsions with low oil concentrations; part (b) for emulsions with higher oil concentrations (note the different vertical scales on (a) and (b)).

Assuming that contributions to light diffraction from the thin oil film that appears at the free surface are small, we convert our results to oil concentration using a Beer-Lambert-type expression derived from first principles [32] (see Methods for further details),

$$-\log_{10}(1 - \hat{I}(t)/\hat{I}_0) = (\epsilon c(t) + b)d(t), \tag{2}$$

where $\hat{I}(t)$ is the time-dependent level of the gray shade of the drop and $\hat{I}_0$ is its initial value. Moreover, $d(t)$ is the time-dependent light path length through the drop, $\varepsilon$ is the light absorbance coefficient, $b$ is a calibration parameter accounting for light reflection from the substrate below the drop (see SI), and $c(t)$ is the concentration of oil droplets in the emulsion. The measurement is conducted in lab ambient conditions and is stopped when the drop loses approximately 20% of its initial 10 μl volume. Beyond this point, the three-phase contact line of the emulsion drop with the solid substrate and vapor tends to move, and the shape of the sessile drop deforms, reducing the measurement precision.

Fig. 10 shows time variations of the oil droplet concentration, $c(t)$, obtained using equation (2) with appropriate values for $\hat{I}(t)$, $\hat{I}_0$, $d(t)$ measured from the experiment, with $\epsilon$, $b$ obtained using calibration (see Methods). We measure the evolution of the oil content under the drop apex for three different initial oil concentrations. In the absence of SAW, we observe that the oil concentration reduces over time for all considered initial oil concentrations. This is attributed to the transfer of oil droplets in the emulsion to an oil film at the emulsion drop surface, a process enhanced by the evaporation of water [22,23]. Since this oil film diffracts light ineffectively compared to the oil droplets in the emulsion, the level of light diffraction is reduced as time progresses.

Once the SAW is applied, the results become more elaborate, with the trend depending on the overall oil concentration. In the case of 50% initial oil concentration, we observe from Fig. 10 that the temporal oil content near the drop apex decreases over time in the presence of SAW. In the case of the initial 30% oil content, the temporal oil content increases over time, and in the case of an initial 40% oil content, the oil content appears to fluctuate over time. These opposing trends in the content of oil near the drop apex are attributed to three factors: (1) the enhanced rate by which oil leaves the drop under SAW excitation [11,12,15,16], reducing the concentration of oil in the drop; (2) the rate at which water leaves the drop by evaporation [22,23], which is also enhanced when increasing the intensity of the SAW; see SI for further details about SAW-enhanced water evaporation; and (3) the spatial variation of oil droplet content due to the SAW 'pushing' the oil droplets to the back of the drop - this effect is enhanced when reducing the concentration of oil therein and increasing the intensity of the SAW.

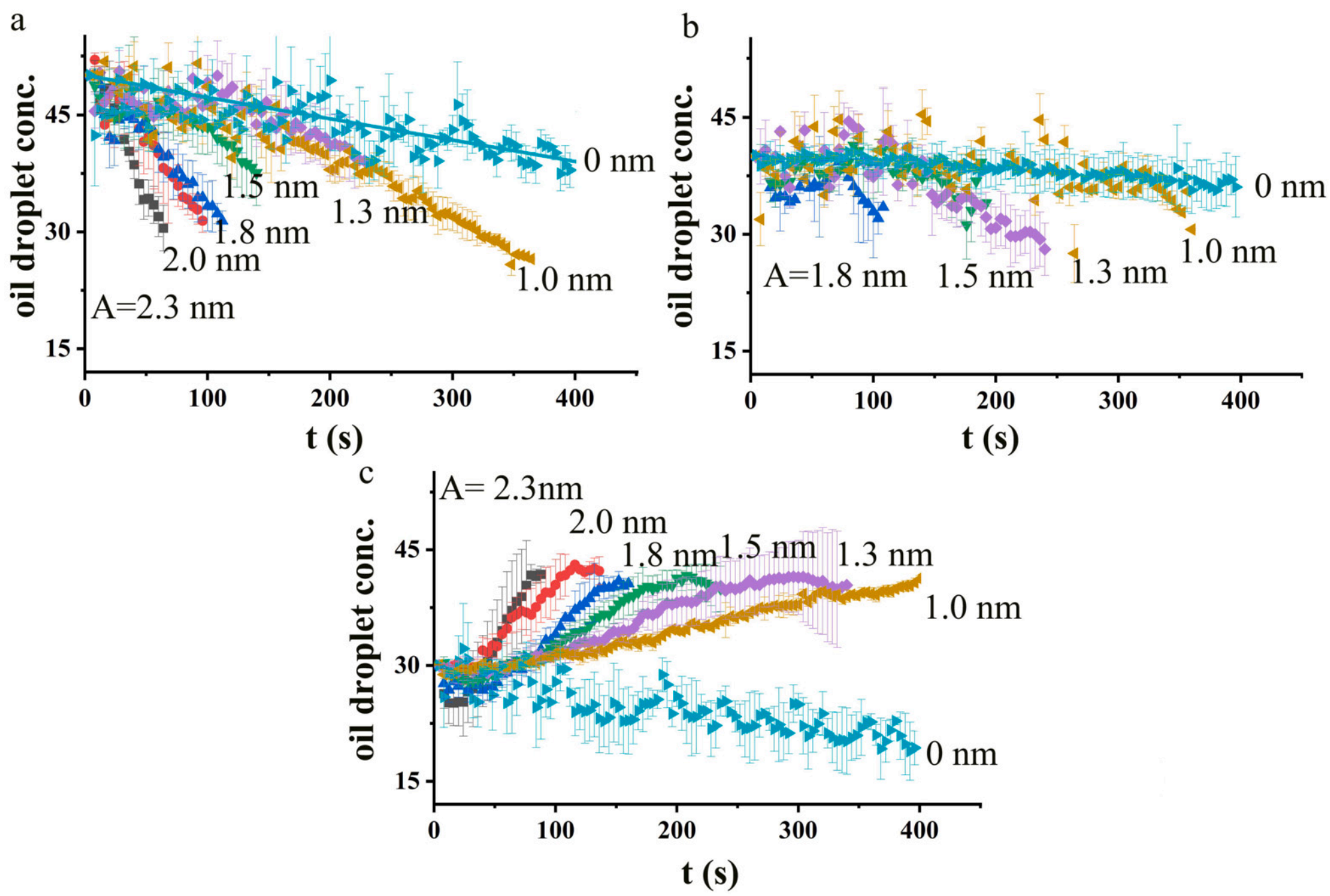

**Fig. 10.** Time ($t$) evolution of oil droplet concentration from the moment SAW was applied to the substrate in different emulsion sessile drops under SAW excitation for initial oil concentration of (a) 50% (b) 40%, and (c) 30%, where the error bars show average deviations in value taken from at least 2 repetitions of the experiment; we give the corresponding SAW intensity $A$ for each experiment on the graphs. The measurement is taken from above through a square of $0.5 \times 0.5$ $mm^2$ at the apex of the drop, while changes in the drop height are accounted for when converting the level of liquid diffraction to oil droplet concentration; the solid lines are guides to the eye where useful.

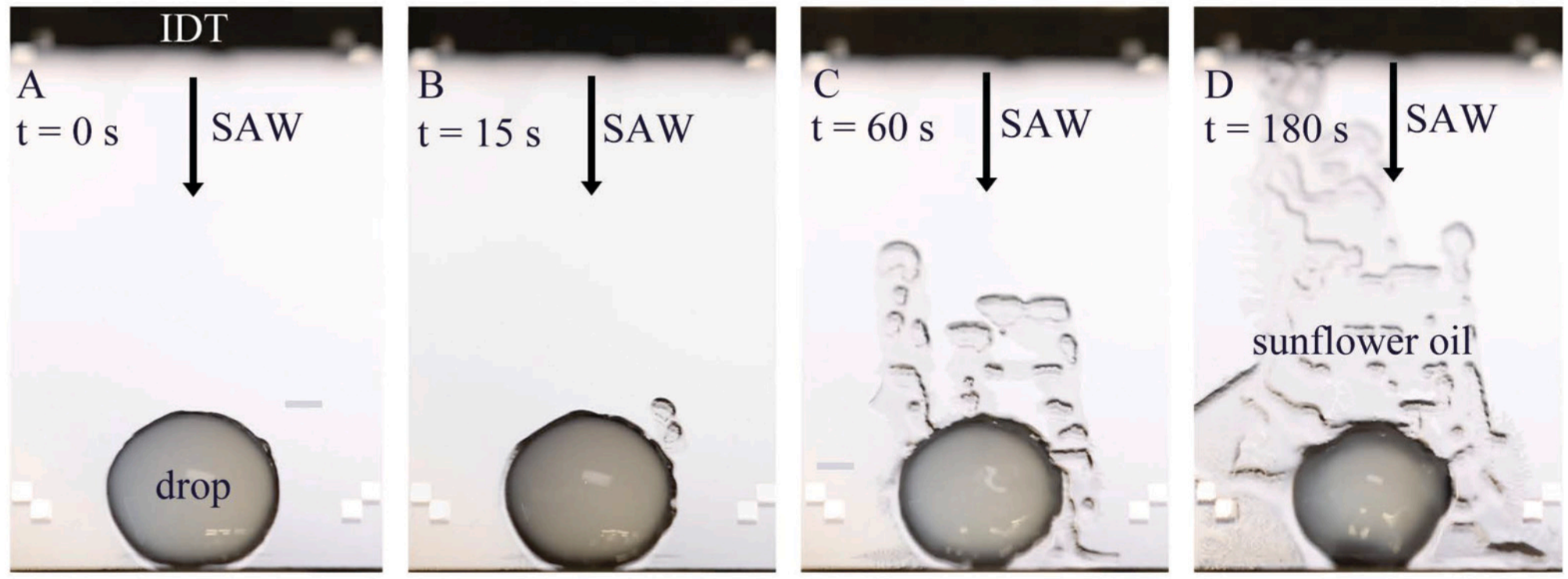

**Fig. 11.** Top view of a 10 μl emulsion drop of sunflower oil under the influence of a SAW surface displacment amplitude ($A$) of 1.2 nm. (A) At $t = 0$ ($t_w = 95$ s after the initial excitation with SAW), we observe the appearance of oil at the rim of the drop. We find that similarly to the case of silicon oil, (B) initially the sunflower oil phase leaves the emulsion drop transverse to the path of the SAW and (c) evolves to become a discontinuous oil film, (D) eventually dynamically wetting the solid surface of the SAW actuator opposite the path of the SAW. See video #6 in Supplementary Information (SI). The emulsion used for this experiment is prepared by following the same procedure as implemented for silicon oil.

One interpretation of the results is that a larger initial oil concentration (Fig. 10a) increases the rate at which oil leaves the drop. An initial oil droplet concentration of 40% appears to introduce a threshold, specific to the conditions in our experiment, for the ratio of the rates at which oil and water leave the drop. At this value, both appear to leave the drop at similar rates, which gives rise to an oil droplet concentration that varies little over time. An initial oil concentration above the 40% threshold results in oil leaving the emulsion drop at a higher rate (leading to negative slope of the data in Fig. 10a), and one below the 40% threshold results in oil leaving the drop at a smaller rate (positive slope of data in Fig. 10b).

This interpretation, however, ignores the internal spatial distribution of oil droplets in the emulsion drop, generated by the SAW, and also that the ratio between the rates at which oil and water leave the drop appears to be determined by the initial oil droplet concentration rather than by its temporal concentration. More research is needed to understand precisely the influence of the interplay between these different mechanisms on the evolution of the oil droplet concentration in the emulsion drop.

## 5. An emulsion of sunflower oil in water

The experiments discussed so far use silicone oil—a model non-organic oil commonly employed in experiments, belonging to a chemical family of oils used for engines. To demonstrate the general application of our work, we repeat our experiment using sunflower oil—an edible organic oil commodity purchased from a nearby supermarket. As detailed in the Methods section, the same procedure (including emulsion preparation with surfactants (SDS) and using SAW to extract oil out of an emulsion drop) is used for the sunflower oil emulsion. The physics observed in the experiment is similar to that observed previously using silicone oil. The surface tension of sunflower oil is ∼ 34 mN/m [33], which is different from the characteristic 20 mN/m surface tension of the silicone oil family. The sunflower oil, similarly to the 20 mN/m silicone oil, satisfies $\theta^3/\mathrm{We} \ll 1$ (based on the oil properties, where the solid/oil/vapor contact angle is small, $\theta \ll 1$) and leaves the emulsion drop under the action of SAW. The water phase, however, satisfies $\theta^3/\mathrm{We} > 1$ (based on the water/surfactant properties, where the solid/water/vapor contact angle is finite, $\theta \approx 30^\circ - 60^\circ$) and remains at rest. In Fig. 11, we see that the sunflower oil films leave a sessile drop of an emulsion comprised of the oil in water.

## 6. Conclusion

Based on prior findings on the spreading of oil [34,35] and water [15–17] films under the action of surface acoustic waves (SAWs) due to their different wettability properties [36], we hypothesize that SAWs may be used to break oil-in-water emulsions. A balance between SAW-induced stress and capillary stress at the free surface of the two liquids supports the dynamic wetting of the solid by the low surface tension oil phase as it leaves the emulsion drop; the high surface tension water phase remains at rest in the drop.

In our experiments, we place sessile 10 μl oil-in-water emulsion drops atop a 20 MHz-frequency SAW actuator. The emulsion drops contain oil droplets within a continuous water phase. The oil droplets must translate to the free surface of the emulsion to be captured by the SAW that facilitates the leakage of the oil phase away from the emulsion drop, while the water phase remains still—our key goal. In addition, we observe an array of flow instabilities and other phenomena in the presence of the SAW. We note that the opposite case of water in oil emulsion drops is much simpler and is not studied here: the SAW actuates the continuous oil phase of the emulsion in a manner similar to observations in previous studies of a pure oil phase [34,35].

The process by which oil leaves the oil-in-water emulsion appears to comprise two parts: In the first part the oil droplets in the emulsion form an oil film atop the free surface of the emulsion drop. This is favorable thermodynamically due to the low surface tension of the oil. Kinetically, this process appears to be aided by the evaporation of the water phase, which enhances the transport of oil droplets to the free surface of the emulsion drop [22,23]. Moreover, the SAW generates acoustic streaming in the drop that introduces heat due to viscous dissipation [37], enhancing the rate of evaporation. In Fig. 5 at the SI, we illustrate the enhancement of water evaporation in the presence of SAW by plotting the time to full evaporation of a 10 μl water/surfactant solution. The drop evaporation time reduces from around 1000 to 100 seconds with increasing the SAW intensity. The second part of the process is the transport of the oil film off the free surface of the emulsion drop and to the solid surface of the SAW actuator by the action of acoustic stress. There, the oil film dynamically wet the solid surface of the actuator.

The acoustic radiation pressure, generated in the emulsion drop by the SAW, spatially redistributes the emulsion droplets therein. The concentration of the emulsion droplets is reduced at the front of the emulsion drop (the portion of the drop that first encounters the traveling SAW in the slid substrate) and increased at its rear. The reduced oil concentration at the front of the drop appears to inhibit the appearance of oil at the front of the emulsion drop due to the local lack of oil droplets. Instead, we observe ‘fingers’ of oil films to initially appear at the sides of the emulsion drop, transverse to the source of the SAW, and at its rear, where the oil droplet concentration appears to increase under the action of the SAW. In particular, oil films appear initially to leave the drop predominantly in the direction transverse to the SAW. At later times, the oil film outside the drop spreads in the direction opposite the SAW propagation, which is compatible with the classic *acoustowetting* phenomenon [34,35]. This is counter-intuitive, since one expects the oil to leave the drop from the side that faces the origin of the SAW, which was observed in previous studies of acoustowetting where only one liquid phase, water or oil, was used. We conclude that the physics of extracting a liquid film from a pure liquid reservoir using SAW differs from the physics of extracting a liquid film from a complex liquid, such as the emulsion drops in our experiment. However, since the oil film appearing at the solid surface is approximately 25 μm thick, which is between 1/4 and 1/2 of the wavelength of the ultrasound that leaks off the SAW into the liquid, it appears that the observed phenomenon is associated with *acoustowetting*, similarly as for pure liquids.

We observe several types of interfacial instabilities, a result of the competition between acoustic and capillary stresses in the oil films that leave the emulsion drop. As noted, the oil initially leaves the drop in the form of ‘fingers’ which later merge into a continuous oil film that develops macro-scale spatial variations in thickness prependicular to the SAW path. The film breaks to thick parts—approximately 25 microns in thickness—and thin parts that are several microns in thickness. The lateral length scale for the thickness variations is approximately 0.5 mm—different from the 200 μm wavelength of the SAW.

Moreover, in the presence of SAW, we observe that the thick oil films support a crystal-like cell pattern observed using interferometric measurements. The cell patterns are characterized by a lateral length scale of tens of microns. They are comprised from spatially-periodic thickness-variations of 0.1 to 0.3 μm. These film thickness variations correspond to a quarter of the red laser wavelength that we use for our measurements, and hence are observed during our interferometric measurements. The cell-pattern appears in the presence of SAW and fades away within seconds once the SAW is turned off.

We have also successfully repeated our experiment using an emulsion of (organic) sunflower oil in water, illustrating that our findings are general and not limited to non-organic oil. The technique envisioned herein is based on discriminating over the different wetting properties of the liquids in an emulsion with the underlying solid substrate through a balance between acoustic and capillary stress and complements current emulsion separation techniques. Moreover, using low-power SAW for oil/water phase separation is compatible with a local small-scale treatment of gray water in domestic housing and small industry, which may avoid or reduce the volume of gray water flow in sewerage systems on their way to large dedicated facilities far from residential areas.

Other modern methods used for breaking oil-in-water emulsions are associated with bulk and surface techniques [38]. Our study is related to the surface-based techniques that employ hydrophobic and hydrophilic membranes with carefully tailored surface properties, which are studied for breaking specific emulsions [39]. However, the use of low-power SAW gives the opportunity to use simple flat surfaces for the oil/water phase separation rather than porous membranes of complex structure.

To establish the use of SAW for oil/water phase separation, future work will require the separation of oil from oil-in-water emulsions in liquid tanks, rather than in drops, considering the manipulation of the emulsion by dedicated design of actuators [40]. Furthermore, the study of the efficiency of the oil-water separation process described in this paper is necessary, as well as a further investigation of the array of observed interfacial instabilities.

## 7. Methods

### 7.1. Emulsion preparation and characterization

Preparing emulsions, we used 50 cSt silicone oil (Sigma-Aldrich, 50cSt, 378356-250ML), Sunflower oil (Borges, Refined Sunflower oil enriched with antioxidants, 1L) and HPLC water (Macron Fine Chemicals, 99.8%, ChromAR@), in addition to Sodium Dodecyl Sulfate, i.e., SDS, (Sigma-Aldrich, ≥99.0%, (GC)74255-250G) and Tween 20 (Sigma-Aldrich, ≤3.0% water P1379-100ML) as surfactants. The emulsions (both silicone oil/sunflower oil emulsion) were prepared by mixing silicone oil and aqueous surfactant solution (4mM for SDS, 0.5%wt for Tween 20. We commenced the preparation by emulsifying mixtures of oil and water/surfactant solutions (total volume of 50 ml) using Rotor-stator mixer (PRO Scientific Inc. Bio-Gen Series PRO 200) for 60 seconds at maximum power. The emulsion droplets were then further reduced in size by sonication, applying 180 seconds of ultrasonic emulsification (Qsonica LLC, Q500) with 30% power output, 5/20 pulse mode. The energy (fixed power & changing time) of mixing/sonication was proportionally reduced based on the volume of liquid. We used Dynamic Light Scattering (DLS) measurement of the emulsions using Zetasizer Ultra (Zetasizer Ultra, Malvern, UK) to measure the size distribution of the oil droplets in the emulsions. The emulsions underwent 100 times dilution using 4 nM SDS solution and where relevant Tween 20 solution to satisfy the dilution requirement of the DLS measurement.

### 7.2. Actuator fabrication and experimental procedure

To fabricate surface acoustic wave (SAW) actuators, we used standard lift-off photolithography to fabricate 24 electrode pairs of 5 nm titanium / 1 μm aluminum interdigitated electrode (IDT) and bus bars for connection to external electric circuit atop 0.5 mm thick, 128° Y-cut, X-propagating, single-crystal lithium niobate piezoelectric wafer (Roditi International, UK) wafers. The wafer was subsequently cut to individual 20 MHz-frequency SAW actuators. We power the actuators using a signal generator (Rohde&Schwarz, SMB100A microwave signal generator) and amplifier (Tabor Electronics, model A10160, Israel) and measured the nanometer normal surface displacement amplitude (intensity) of the Rayleigh SAW in the actuators using a scanning laser Doppler vibrometer (MSA-500, Polytech). This measurement further confirmed that the SAWs are traveling waves. Before every experiment, we washed the SAW actuator for 30 s with 4 different solvents in the following order: acetone (AR-b, 99.8%, 67-641, Bio-Lab Ltd.), 2-propanol (AR-b, 99.8%, 67-63-0, Bio-Lab Ltd.), ethanol (CP-p, 96%, 64-17-5, Bio-Lab Ltd.), and water (ChromAR@ HPLC, Macron Fine Chemicals).

We fix and connect the SAW actuators by pogo-pins (BC201403AD, Interconnect Devices, Inc.) attached to a 3D-printed elastomeric stage (shown in Fig. 1(a)) under ambient lab environment ($20 \pm 2°$ Celsius and $50 \pm 5\%$ humidity) and in a humidity chamber ($20 \pm 2°$ Celsius and $85 \pm 5\%$ humidity). To fabricate the humidity chamber shown in Fig. 1(c), we employ a commercial glass petri-dish and a plastic cover. We further put a small hole in the plastic cover to connect the SAW actuator to power via electric wires. The wires were attached to the SAW actuator using a conductive silver-epoxy (EPO-TEK@ H20E, Epoxy Technology, Inc.). Another hole was used to monitor the internal temperature and humidity in the chamber using a dedicated sensor (EARU Electric). Moreover, the experiments are recorded using a camera (EOS R5, Canon) fit with a macro lens (RF 100 mm F2.8L MACRO IS USM, Canon). We further employ a light diffuser (a rough plastic surface) between the light source and the experiment to reduce the level of localized concentrated light areas. We capture videos of the experiment using camera Aperture 4.0, Shutter Speed 1/320, and ISO (5000) to reduce the effect of the camera light auto-adjustment.

During the experiment, we place 10 μl emulsion drops atop the SAW actuator, far from the electrodes, see Fig. 1. The oil droplets in the emulsion have an average diameter of 230 nm (the diameter of droplets is calculated based on the number of drops in the emulsion), with oil concentration of 10-50% by volume. The emulsions are stabilized using surfactants (SDS or TWEEN20). The SAW actuator is activated immediately following the placement of a drop. We measure the temperature of the solid surface near the emulsion before and after the activation of SAW. When using the humidity chamber, we waited for 10 minutes from the moment we placed the drop atop the SAW device and closed the chamber, before applying power to the SAW actuator, to stabilize the humidity therein prior to the commencement of the experiment.

### 7.3. Interferometric oil thickness measurements

We generate FECO fringes to estimate the height of the oil film induced by the Surface Acoustic Wave (SAW). For the light inteferometry experiments, we mount a level-1 laser generator (635 nm, LM-6305MR, Lanics) and a camera (EOS R5, Canon) at similar angle with respect to the horizon to observe the FECO patterns induced by the light interferometry. The angles are monitored using a digital protractor attached to the laser generator. During the measurement, we shine the 635 nm laser at the oil film to result in constructive and destructive light interference patterns (light fringes) at the oil free surface due to the additional optical path traveled by the light reflected from both the oil/air and oil/solid interfaces.

Regarding the uncertainty in our measurement, four independent measurements give less than 3% deviation between the different measured values. The measurements given are taken 1 second after the SAW in the substrate is turned off, following the disappearance of the cell-like interferometric patterns atop the films. To estimate the uncertainty in the measured film thickness due to the 1 second waiting time, we resort to the thin film equation [41]: accounting for mass and momentum conservation, the long wave approximation of film dynamics, which is governed by capillary stress and viscous dissipation, is given by the film equation $\partial h/\partial t = (\gamma/3\mu)\nabla_s \cdot \left(h^3 \nabla_s \nabla_s^2 h\right)$, where $h$ and $t$ are the local film thickness and time, $\gamma = 20 \times 10^{-3}$ N/m and $\mu = 50 \times 10^{-3}$ Pa· s are the silicone oil film surface tension (against vapor) and viscosity in our experiment, and $\nabla_s$ is a gradient operator along the solid surface. The film thickness scales like its measured value, $h \approx h_{\text{max}} = 25$ μm. The gradient operation scales like the inverse of the characteristic lateral length scale of the film, i.e., $\nabla_s \approx 1/l_{\text{film}} = 1\ \text{mm}^{-1}$. Hence, the scale for the rate of change in film thickness is given by $\partial h/\partial t \approx (\gamma/\mu) \times h_{\text{max}}^4/l_{\text{film}}^4 \approx 10^{-1}$ μm/s. The characteristic change in local film thickness within 1 second is then 0.1 μm. Hence the uncertainty in the maximum film thickness due to the 1 second waiting time in our experiment is $0.1\ \mu\text{m}/25\ \mu\text{m} \approx 0.4\%$. Therefore, the overall uncertainty in the maximum measured film thickness is approximately $3.4\% \approx 4\%$, when accounting for both variability between measurements, and the relaxation of the oil film in the 1 second time waiting time from the moment we turn off the SAW and until we take the measurement.

The change in local film thickness during the one second waiting time is of similar magnitude to the characteristic difference between bright and dark interferometric light patterns of approximately 0.16 μm (a quarter of the laser wavelength). Hence, the micro-cell pattern appearing on top of the macro-film structure as an interferometric pattern in the presence of the SAW is associated with a local film thickness variations of a similar magnitude.

### 7.4. Measuring oil content using a Beer-Lambert-like rule

The simple form of the Bouguer-Beer-Lambert Law gives the equation

$$-\log_{10}(I/I_0) = \varepsilon c d, \tag{3}$$

where $I_0$ is the original light intensity, $I$ is the light intensity after traveling through a medium that contains a molar concentration of objects, $c$, that are of molar absorption coefficient $\varepsilon$, assumed constant throughout each experiment, and $d$ is the path length of the light in the medium. The right hand side of the equation is the light absorbance of the medium, given by $\varepsilon c d$. Moreover, in each experiment, we calibrate (scale) the value of light intensity $I$ and $I_0$ against variations in ambient light by placing the same white paper next to the experiment. As a precaution, the values of $I$ and $I_0$ are normalized against (divided by) the color intensity of the white paper in each experiment, so that we are able to compare between different experiments, taken at different times. The corresponding normalized values are denoted by $\hat{I}$ and $\hat{I}_0$, respectively.

The brightness that we observe on the emulsion drop is a result of light scatter by the emulsion. The emulsion drop becomes continuously darker (and more transparent) when the emulsion concentration is reduced and when the thickness (height) of the drop is reduced. Monitoring the change in the light intensity (brightness) of the emulsion over time, $\hat{I} = \hat{I}_0 - \hat{I}(t)$, we rewrite (3) in the form

$$-\log_{10}((\hat{I}_0 - \hat{I}(t))/\hat{I}_0) = -\log_{10}(1 - \hat{I}(t)/\hat{I}_0) = (\epsilon c(t) + b)d(t), \qquad (4)$$

where we added the constant $b$ to account for light reflection from the substrate below the drop. We find the constants $\epsilon$ and $b$ using a calibration curve constructed from measuring a series of emulsion drops of known oil concentration. The magnitude of $d(t)$ is twice the drop maximum thickness (height) and is measured over time, $t$, throughout each experiment in tandem with the light intensity measurement. Following the experiment, we convert the measured light intensity at the apex of the drop to an approximation of the time-dependent concentration of oil droplets therein, $c(t)$. Moreover, since in our experiments the SAW renders a spatial redistribution of oil droplets inside the bulk emulsion drops, the measured quantity $c(t)$ is an approximation of the average concentration of emulsion droplets in the drops.

### CRediT authorship contribution statement

**Yifan Li:** Writing – original draft, Formal analysis. **Jesús. M. Marcos:** Formal analysis. **Mark Fasano:** Formal analysis. **Javier Diez:** Formal analysis. **Linda J. Cummings:** Writing – original draft, Formal analysis. **Lou Kondic:** Writing – original draft, Investigation, Funding acquisition, Formal analysis. **Ofer Manor:** Writing – original draft, Methodology, Investigation, Funding acquisition, Formal analysis, Conceptualization.

### Declaration of competing interest

The authors declare the following financial interests/personal relationships which may be considered as potential competing interests: Ofer Manor, Lou Kondic reports financial support was provided by US-Israel Binational Science Foundation. Ofer Manor, Lou Kondic reports financial support was provided by ACS Petroleum Research Fund. J. M. Marcos reports financial support was provided by Spanish Ministerio de Universidades. If there are other authors, they declare that they have no known competing financial interests or personal relationships that could have appeared to influence the work reported in this paper.

### Acknowledgement

This work was supported by US-Israel Binational Science Foundation (BSF) under grant No. 2020174 and by the donors of ACS Petroleum Research Fund under Grant PRF# 62062-ND9. J.M.M. is grateful to the Spanish Ministerio de Universidades for a predoctoral fellowship No. FPU2021-01334. J.A.D. acknowledges support from Consejo Nacional de Investigaciones Científicas y Técnicas (CONICET, Argentina) with Grant PIP 02114-CO/2021 and Agencia Nacional de Promoción Científica y Tecnológica (ANPCyT, Argentina) with Grant PICT 02119/2020.

### Appendix A. Supplementary material

Supplementary material related to this article can be found online at https://doi.org/10.1016/j.jcis.2025.138442.

### Data availability

Data will be made available on request.